# DIFET: DISTRIBUTED FEATURE EXTRACTION TOOL FOR HIGH SPATIAL RESOLUTION REMOTE SENSING IMAGES

S. Eken [a,*], E. Aydın [a], A. Sayar [a]

[a] Kocaeli University, Dept. of Computer Engineering, 41380, İzmit Kocaeli, Turkey – (suleyman.eken@kocaeli.edu.tr, eray.aydin@live.com, ahmet.sayar@kocaeli.edu.tr)

**Commission VI, WG VI/4**

**ABSTRACT:**

In this paper, we propose distributed feature extraction tool from high spatial resolution remote sensing images. Tool is based on Apache Hadoop framework and Hadoop Image Processing Interface. Two corner detection (Harris and Shi-Tomasi) algorithms and five feature descriptors (SIFT, SURF, FAST, BRIEF, and ORB) are considered. Robustness of the tool in the task of feature extraction from LandSat-8 imageries are evaluated in terms of horizontal scalability.

## 1. INTRODUCTION

Acquiring remote sensing data has been improved to an anomalous line. The volume of global data archive could even be on the Exabyte level because of both of improvements in spatial, spectral, radiometric, and temporal resolutions and increasing number of satellites by year by. As a result, we have big remote sensing data with characteristics of multi-source, multi-scale, high-dimensional, and etc.

The more characteristics, the more information we obtain. However, there is no doubt that most of existing techniques and methods are too limited to solve all the problems of remote sensing big data due to its complexity. Also, processing remote sensing data is time consuming, especially when working with such high resolution data. Since almost all algorithms and models have to consider the intrinsic and extrinsic characteristics of data, applications now have to adapt to the great changes from remote sensing big data.

Feature extraction is one of the most essential steps in remote sensing for different application areas such as object detection (Sayar *et al.*, 2014; Eken and Sayar, 2015), target tracking (Meng and Kerekes, 2012), image matching (Wang *et al.*, 2012; Ling *et al.*, 2016), image stitching (Sayar *et al.*, 2013), and etc. The extracted features can be classified into two main categories: global image features (GIFs) and local image features (LIFs) (Dimitri *et al.*, 2005). Aforementioned applications tend to use either GIFs or LIFs. While GIFs (color histograms, principle component analysis, and etc.) describe an image as a whole, LIFs represent image patches. GIFs have the ability to generalize an entire object with a single vector. So, GIFs are not capable of matching local regions which are prominent to the object or scene in the image. LIFs are computed at multiple points in the image and are consequently more robust to occlusion, clutter and illumination change. Also, they are invariant to translation, rotation, scale, affine transformation, and presence of noise, blur etc. LIFs also are divided into two classes: line- and region-features and point features. As former one is more difficult and less accurate, the point-based methods are much more widely used. Also, LIFs have good locality, they do not require the global communications between LIFs process. In this paper, we focus on LIFs and implementation of distributed feature extraction tool (DIFET) for high spatial resolution remote sensing images. To extract LIFs features from high spatial resolution remote sensing images, we use Hadoop Image Processing Interface (HIPI)1, which is based on MapReduce approach. Apache Hadoop is placed at the core of the framework. To realize such a system, we first define a mapper function, and its input and output formats. Then, the scalability analysis is performed.

The remainder of this paper is organized as follows. Section 2 firstly presents literature review on distributed image processing tools and frameworks then gives point and line-and-region features and descriptors implemented in DIFET. Section 3 explains DIFET architecture in detail. Experimental setup and performance results are given in Section 4. Section 5 concludes the paper and describes the directions for future work.

## 2. OVERVIEW ON FEATURE EXTRACTION

### 2.1 Distributed Processing Tools and Frameworks

When a huge size of high spatial resolution remote sensing images struggle for extracting features, it makes efficient access to the images and management of potentially heterogeneous system resources for data processing a time consuming task. By using desktop based sequential systems, feature extraction for huge sized data takes hours. In this work, we describe a tool to execute distributed algorithms on a Linux cluster using Hadoop MapReduce (Dean and Ghemawat, 2008). Recently, the MapReduce framework has become the de facto standard for handling large scale data processing tasks, and it has many salient features such as massive scalability, fault-tolerance, easy programmability and low deployment cost. With the success of high performance computing (HPC) technology and MapReduce paradigm, a number distributed and parallel computing based techniques have been proposed to enable large scale remote sensing image processing on large datasets in the literature. Some of them are listed as following paragraph.

Zhanfeng *et al.* (2007) developed a distributed processing system enabling image segmentation, image classification, image target recognition, and etc. for processing remotely sensed images. Ariel *et al.* (2009) proposed a MapReduce model to solve two spatial problems: bulk-construction of R-

---

Trees and aerial image quality computation on a Google & IBM cluster. Golpayegani and Halem (2009) suggested parallel computing framework for satellite data processing. Zhenhua *et al.* (2010) introduced parallel k-means clustering of remote sensing images based on MapReduce programming model. Junfeng *et al.* (2012) designed a remote sensing image service framework to providing static and dynamic web map service. Mamta *et al.* (2013) reviewed recent development in high performance computing (HPC) technology for satellite data processing. In the literature, there are many works and applications related to or including feature extraction phase. However, to our knowledge no previous feature extraction tool or work has been done in parallel or distributed manner in big data concept for massive sized remote sensing data.

**2.2 Extraction of Local Image Features**

In the computer vision based applications, it is very important to find specific patterns or specific features which are unique, which can be easily tracked, which can be easily compared. Figure 1 shows the importance of feature detection and description in better and simpler way. Blue patch is flat area and difficult to find and track. Wherever you move the blue patch, it looks the same. For black patch, it is an edge. If you move it in vertical direction (i.e. along the gradient) it changes. Put along the edge (parallel to edge), it looks the same. And for red patch, it is a corner. Wherever you move the patch, it looks different, means it is unique. So basically, corners are considered to be good features in an image.

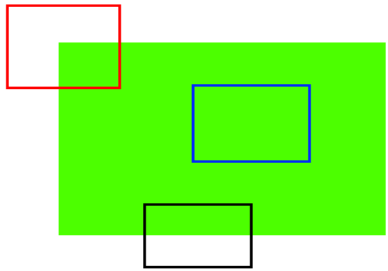

Figure 1. Understanding features

Good features can be found by looking for the regions in images which have maximum variation when moved in all regions around it. Also, finding these image features is called Feature Detection (FDet). After finding the features in image, same ones can be found in the other images. The best way to do is describing the region around the feature so that it can find it in other images. So called description is called Feature Description (FDes). With the features and its description, same features can be found in all images and aligned them, stitched them or done whatever we want. The extraction of features from an image is a job that can return different results, depending on the used methods. In consideration of speed and accuracy of finding this features there can distinguish many different algorithms, each of them has strengths and weaknesses. At following three sub-sections we are going to describe point/corner detectors, line-and-region detectors and discriptors, respectively. We have restricted this work to approaches carried out by DIFET.

**2.2.1 Point (corner or interest point) detectors**

A corner or an interest point is a point in an image which has a well-defined position and can be robustly detected. In practice, most so-called corner detection methods detect interest points in general. Interest-point detector can detect: (i) points on corners and (ii) points on blob like structures. Many corner detection algorithms have been proposed by the researchers (Trujillo and Gustavo, 2008). In this paper, we focus on Harris corner detection and Shi-Tomasi corner detector. Features from accelerated segment test (FAST) is also used to extract feature point. The most promising advantage of the FAST corner detector is its computational efficiency (Rosten and Drummond, 2006).

**2.2.2 Line (edges) and region (blob) detectors**

An edge can be defined as a location of rapid intensity change. Edge detection approaches can be divided into two classes: (i) search-based (or gradient based) and (ii) zero-crossing based (or Laplacian based). The search-based methods detect edges by first computing a measure of edge strength and then searching for local directional maxima of the gradient magnitude (Prewitt, 1970; Roberts, 1965) The zero-crossing based methods search for zero crossings in a second-order derivative expression.

Blob detectors can be used to provide complementary information about regions, which is not obtained from edge detectors or corner detectors. Blob detection approaches can roughly be grouped into the following categories: (i) template matching, (ii) watershed detection, (iii) blob detection through scale-space analysis, and (iv) color tensor analysis. Detailed explanations can be found in (Lindeberg, 1994; Ming and Ma, 2007).

**2.2.3 Feature Descriptors**

After detecting interest points/features, it is need to describe them for recognizing them later. In many cases, the local appearance of features will change in orientation and scale, and sometimes even undergo affine deformations. So, image descriptors must be invariant to such changes. At following paragraphs, a few of these descriptors are described in more detail.

The scale-invariant feature transform (SIFT) feature detection algorithm is developed and pioneered by David Lowe. SIFT is a four stage process that creates unique and highly descriptive features from an image and enables finding correspondence between parts of images (Lowe, 2004). Speeded up robust features (SURF) s partly inspired by SIFT descriptor (Bay *et al.*, 2008). To detect interest points, SURF uses an integer approximation of the determinant of Hessian blob detector. Calonder *et al.* (2010) propose to use binary strings as an efficient feature point descriptor called BRIEF (Binary Robust Independent Elementary Features). BRIEF directly builds short descriptors by comparing the intensities of pairs of points without ever creating a long one. Ethan *et al.* (2011) propose a fast robust local feature detector named Oriented FAST and rotated BRIEF (ORB). It is based on the FAST feature detector and the visual descriptor BRIEF. Its aim is to provide a fast and efficient alternative to SIFT.

Feature extraction procedure can be done in parallel manner with two ways. One way is usage of special hardware such as GPUs, FPGA, and etc. Other way is usage of software based approaches. DIFET is in latter category. Next section explains proposed architecture.

## 3. DIFET ARCHITECTURE

Apache Hadoop is an open source distributed master-slave framework. It consists of two main parts: (i) scalable and reliable file system named Hadoop Distributed File System (HDFS) for storage and (ii) distributed processing part named MapReduce for computational capabilities. There are two general classes of nodes involved in Hadoop. These are master nodes called namenodes, and slave nodes called datanodes. The namenode is a kind of manager keeping track of both actions of datanodes and metadata for all directories and files. A Job in Hadoop is run in MapReduce approach and in parallel. A job in MapReduce contains three phases: map, shuffle, and reduce. To get an expected performance gain by running a job on Hadoop, map and reduce phases of a job need to be defined very carefully. For more details about the framework, its open-source implementation can be found in (White, 2015).

DIFET architecture utilizes Hadoop Image Processing Interface (HIPI), which is based on MapReduce approach. HIPI facilitates efficient and high-throughput image processing with MapReduce style parallel programs typically executed on comodity nodes. HIPI creates HipiImageBundles (HIB) in Hadoop Distributed File System (HDFS) storage which stores collection of images in a single file with some meta data information. HIB bundle is the primary input of an HIPI program which work on map-reduce framework. HIPI also provides integration with OpenCV. In the literature, there are a little works using HIPI. Wilder *et al.* (2015) extend HIPI to handle images in TIFF or GeoTIFF format since HIPI does not support all image formats. Various image processing operations such as filters, variance, clustering or dimensionality reduction have been tested on LandSat satellite images. Basil *et al.* (2015) propose a surgical video analysis system that uses Hadoop to describe the surgical instruments used in very large-scale surgical surgery videos. Akkoyunlu *et al.* (2016) conduct a performance study on the detection of biometrics belonging to face regions over two different datasets. Changes have been made to the HIPI interface's FloatImage, HipiImageHeader, and ImageCodec classes in this work.

General architecture for feature extraction is illustrated in Figure 2.

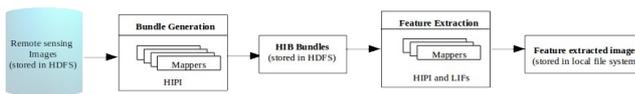

Figure 2. Proposed architecture

Hadoop framework stores satellite images in HDFS and the data is distributed among several datanodes. HIPI uses HIB bundles to stores images so that each mapper is provided with a single image. In feature extraction phase, local features are extracted from satellite images in the HIB bundles. HIPI interface allow each image in HIB bundle to be processed by individual mappers. Each mapper extracts local features and descriptors using aforementioned algorithms in Section II. To better understanding the mapper functions of algorithms, pseudo-codes for Harris corner detector and SURF descriptors are given as following. In DIFET architecture, uses HibInputFormat, which receives the HIB file and provides the HIPIImageHeader and the image (in FloatImage format) as key and value pair respectively to the mapper. In-turn the mapper converts it to OpenCVMatWritable format and then processes it by feature extraction/descriptor algorithms and thereafter saving the Mat image onto the HDFS which is a one time process.

Map function for Harris detector
1. Convert FloatImage to OpenCv matrix
2. Convert image to grayscale
3. Apply Harris corner detection to grayscale image
4. Convert the matrix to FloatImage.
5. Save FloatImage to hdfs with jpeg encoder

Map function for SURF descriptor
1. Convert FloatImage to OpenCv matrix
2. Convert image to grayscale
3. Set surf hessian threshold to 400
4. Apply SURF algorithm to grayscale image
5. Draw obtained keypoints to the matrix
6. Convert the matrix to FloatImage.
7. Save FloatImage to hdfs with jpeg encoder

## 4. EXPERIMENTAL RESULTS

In this section, we present the experimental evaluation of distributed local feature extraction algorithms for high spatial resolution remote sensing images. The images in this study were taken from a recently-launched LandSat-8 satellite and had a resolution of around (7000x7000). These images are formatted in RBGA color, meaning that each pixel in an image occupies a 32-bit size. A typical example of such an image with a size of 7681x7831 allocating 230 MB (32x7681x7831 bits) in the memory.

All experimental results are obtained using two nodes, four commodity machines (multi node/cluster) and one node differently to show scale-out behavior of algorithms. Each machine has a single quad-core Intel Core i7-950 3.0 GHz processor, 8 GB DRAM memory, and two 1 TB SATA2 7200RPM hard disks. The operating system is Ubuntu Linux 10.10. All nodes are interconnected using Ethernet switch. Apache's Hadoop version 1.02 is installed for MapReduce platform. One of nodes is configured as both the jobtracker and namenode, while the rest of the compute nodes are configured as task trackers and datanodes.

Table 1 shows the horizontal scalability analysis of algorithms and Table 2 represents number of features obtained from algorithms.

| Alg. | Running times (sec) | | | | | |
|---|---|---|---|---|---|---|
| | *One node (Matlab)* | | *Two machines (MapReduce)* | | *Four machines (MapReduce)* | |
| # of images | N=3 | N=20 | N=3 | N=20 | N=3 | N=20 |
| Harris Corner Detection | 68 | 600 | 44 | 523 | 24 | 174 |
| Shi-Tomasi | 77 | 441 | 31 | 256 | 10 | 85 |
| SIFT | 4140 | 27981 | 1309 | 8818 | 459 | 2945 |
| SURF | 94 | 546 | 110 | 793 | 39 | 260 |
| FAST | 14 | 95 | 21 | 138 | 6 | 43 |
| BRIEF | 143 | 846 | 86 | 511 | 35 | 316 |
| ORB | 30 | 205 | 26 | 169 | 9 | 58 |

Table 1. Running times of algorithms

| Algorithms | # of images | |
|---|---|---|
| | N=3 | N=20 |
| Harris Corner Detection | 140702 | 943159 |
| Shi-Tomasi Corner Detection | 1200 | 8000 |
| SIFT | 123960 | 832604 |
| SURF | 58692 | 398289 |
| FAST | 707264 | 4762222 |
| BRIEF | 3478 | 23547 |
| ORB | 1500 | 10000 |

Table 2. Number of points

## 5. CONCLUSION

We introduced distributed feature extraction and description tool for high spatial resolution remote sensing images. The proposed tool is based on Apache Hadoop with HIPI. Some well-known feature extraction algorithms and feature descriptors are implemented. Scalability analysis of all approaches show that an increase in working times are observed with an increase in the number of image. Also, the running time on a single machine is more than the distributed architecture.

In summary, our key contribution of this work is as follow:

- Developing remote sensing big data processing tool to extract features,

- Enabling vision task with remote sensing big data such as corner, line, and blob detection and description of features.

### ACKNOWLEDGEMENTS

This work has been supported by the TUBITAK under grant 215E189.